\documentstyle[11pt,newpasp,twoside,epsf,graphics]{article}
\def\edcomment#1{\iffalse\marginpar{\raggedright\sl#1\/}\else\relax\fi}
\reversemarginpar

\begin{document}
\title{A radio-microlensing caustic crossing in B1600+434?}
\author{L.V.E. Koopmans$^{1,4}$, A.G. de Bruyn$^{2,1}$,  
J. Wambsganss$^{3,5}$ \& C.D.~Fassnacht$^6$}

\footnotetext[1]{
Kapteyn Astronomical Institute, P.O.Box 800, 
NL-9700 AV Groningen, The Netherlands;
$^2$NFRA-ASTRON, P.O.Box 2, NL-7990 AA Dwingeloo, 
The Netherlands;
$^3$University of Potsdam, Institute for Physics,
Am Neuen Palais 10, 14469 Potsdam, Germany;
$^4$Jodrell Bank Observ., Lower Withington, 
Macclesfield, Cheshire SK11 9DL, UK;
$^5$Max-Planck-Institut f\"ur Gravitationsphysik,
``Albert-Einstein-Institut", Am M\"uhlenberg 1, 
14476 Golm, Germany;
$^6$NRAO, P.O.Box 0, Socorro, NM 87801, USA}

\begin{abstract}
First, we review the current status of the detection of strong
`external' variability in the CLASS gravitational B1600+434, focusing
on the 1998 VLA 8.5--GHz and 1998/9 WSRT multi-frequency
observations. We show that this data can best be explained in terms of
{\it radio-microlensing}.  We then proceed to show some preliminary
results from our new multi-frequency VLA monitoring program, in
particular the detection of a strong feature ($\sim$30\%) in the light
curve of the lensed image which passes predominantly through the
dark-matter halo of the lens galaxy. We tentatively interpret this
event, which lasted for several weeks, as a {\it radio-microlensing
caustic crossing}, i.e. the superluminal motion of a $\mu$as-scale
jet-component in the lensed source over a single caustic in the
magnification pattern, that has been created by massive compact
objects along the line-of-sight to the lensed image.
\end{abstract}

\vspace{-0.5cm}

\section{Introduction}

Optical microlensing has unambiguously been detected both in our own
Galaxy (e.g. the EROS, MACHO and OGLE collaborations; see these
proceedings), as well as in external galaxies (e.g. Q2237+0305; Irwin
et al. 1989). However, no convincing detections of microlensing in
other wavebands have been claimed thus far. In our Galaxy this is
mostly due to the very low surface number density of bright sources,
other than stars, that are compact enough (i.e. $\la$few mas) to be
significantly microlensed. For cosmological microlensing there are
similar arguments. Because the angular Einstein radius of a point-mass
scales approximately as $\propto$$D^{-1/2}$, where $D$ is the distance
to the lens, the lensed source must be $\la$few\,$\mu$as in angular
size to by microlensed by compact stellar-mass lenses at cosmological
distances (few Gpc). It was thought until recently that only the
optical to $\gamma$-ray emitting regions of quasars and AGNs had these
extremely small angular scales and would still be observable over
cosmological distances.

In this proceeding we will show, however, that `radio-microlensing'
has most likely been detected in the CLASS gravitational lens
B1600+434 (Jackson et al. 1995) and that it promises to be a new and
exciting technique for the study of massive compact objects in
galaxies at intermediate redshifts.

\section{Macro \& microlensing of flat-spectrum radio sources}

The JVAS/CLASS gravitational-lens survey (e.g. Browne et al. 1997) has
discovered at least 17 radio-bright gravitational lens systems. All
systems were selected to have a flat spectrum ($\alpha$$<$0.5 with
$S_{\nu}$$\propto$$\nu^{-\alpha}$) between 5 and 1.4\,GHz. This
ensures that most of these sources are dominated by a compact
radio-bright core. These sources are often variable on short time
scales and in many cases show superluminal motion (e.g. Vermeulen \&
Cohen 1994).

Strong variability seems to imply that these radio sources are very
compact. Based on a simple light-travel-time argument, one expects a
source at a cosmological distance (few Gpc) to have an angular scale
of only a few $\mu$as if it varies significantly on a time scale of
say one month. However, superluminal motion with a high
Doppler-boosting factor ($\cal D$) complicates this argument, because
the intrinsic variability time scale is reduced by a factor ${\cal
D}^{3}$, whereas the intrinsic angular scale of the source is reduced
by only ${\cal D}^{1/2}$ (e.g. L\"ahteenm\"aki, Valtaoja \& Wiik
1999).  Based on variability arguments, the angular size of the source
could therefore be severely underestimated if ${\cal D}\gg 1$.
Furthermore, based on Rayleigh--Jean's law and a typical surface
brightness temperature of $10^{11-12}$\,K (e.g. L\"ahteenm\"aki et
al. 1999), one expects an angular size as much as several tens of
$\mu$as for those sources that have been observed in the JVAS/CLASS
survey. This is much larger than the typical Einstein radius of a
solar-mass object. Overall, it appears that microlensing of these
sources is at most marginal, especially for the brighter sources (tens
of mJy).

However, matters are not as bad as they seem! First of all, lensed
sources are often magnified by the lensing potential. Their intrinsic
flux-density is therefore less than the observed flux-density and,
because surface brightness (temperature) is conserved in lensing, also
their intrinsic angular source size is smaller. Second, flat-spectrum
sources at high redshifts often exhibit jet-structures with
superluminal motion (e.g. Vermeulen \& Cohen 1994). If these jets
contain many distinct `bullets' or shock-fronts, their angular sizes
can be very small, especially if they are significantly
Doppler-boosted (${\cal D}\gg 1$). We expect this to be the case near
the core of these flat-spectrum sources, although components further
along the jet might grow in size and have lower velocities. Hence,
even though the integrated flux-density of the source might be large,
suggesting a large angular size, in reality the likely presence of
compact substructure with high Doppler-boostings makes this argument
significantly weaker.  The expected angular size of these
flat-spectrum subcomponents is
$$\Delta\theta\sim\sqrt{\frac{S_{\rm mJy}(1+z_{\rm s})}{ \mu\,{\cal
D}\,T_{12}}} \left(\frac{\lambda} {1\,{\rm cm}}\right)~~~\mu{\rm
as},$$ where $T_{12}$ is the surface brightness temperature of the
component in units of $10^{12}$\,K, $\mu$ is the magnification by the
lens and $z_{\rm s}$ is the source redshift. For subcomponent
flux-densities ($S_{\rm mJy}$) of a few mJy, an angular size of a few
$\mu$as can be expected. This is small enough to be appreciably
microlensed!

Below we will illustrate this by the CLASS gravitational lens system
B1600+434, in which believe to have detected radio-microlensing of
precisely this type of $\mu$as-scale jet-components (Koopmans \& de
Bruyn 2000; KdB00 hereafter).

\begin{figure}[t!]
\begin{center}
  \leavevmode
\vbox{%
  \epsfxsize=\hsize
  \epsffile{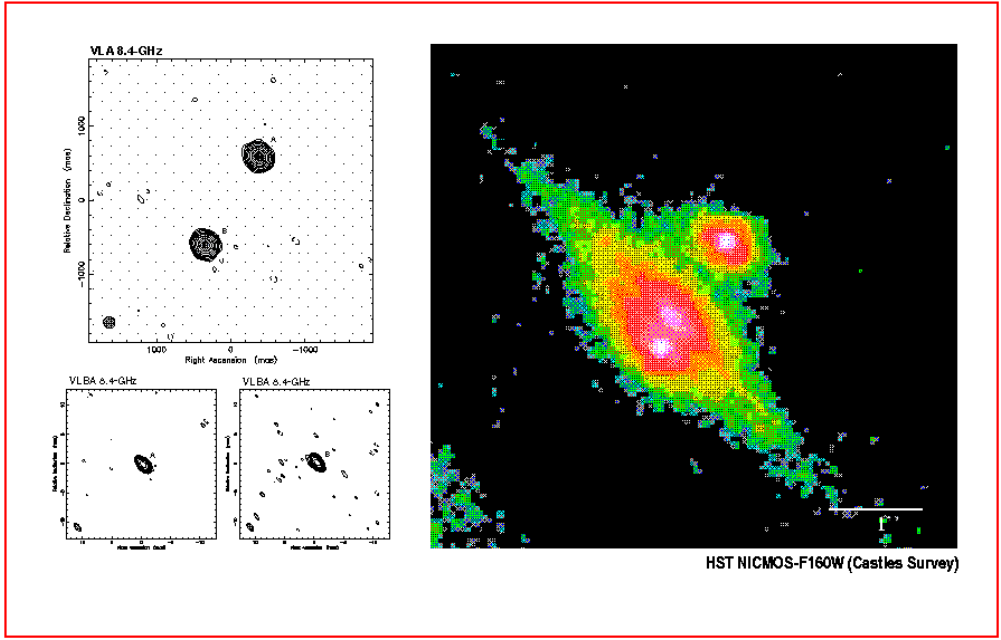}
}
\end{center}
\caption{{\bf Left:} upper -- VLA A-array 8.5-GHz image of B1600+434,
showing the two compact flat-spectrum components, separated by 1.4
arcsec.  lower -- VLBA 8.5-GHz images of the two components, showing
that they are compact down to the milli-arcsec level.  {\bf Right:}
HST NICMOS H--band image of B1600+434 from the CASTLES survey,
showing the edge-on spiral lens galaxy and the two compact
optical components (A and B), corresponding to the
two radio images. }
\end{figure}

\section{CLASS B1600+434}

The CLASS gravitational lens B1600+434 consists of two compact radio
components separated by 1.4 arcsec (Fig.1). The lens galaxy is an
edge-on spiral at a redshift of 0.41, whereas the source has a
redshift of 1.59 (Koopmans, de Bruyn \& Jackson 1998). Image A passes
predominantly through the dark-matter halo of the lens galaxy
($\sim$6\,kpc above the plane of the galaxy), whereas image B passes
through its stellar component (i.e. disk/bulge). The extended optical
emission around image A is thought to be associated with its host
galaxy at $z$=1.59. Both radio components are compact ($\la$1\,mas)
and have a flat radio spectrum. The source is highly variable at
frequencies between 1.4 and 8.5\,GHz (Koopmans et al. 1998).

To determine the time delay between the lensed images, we monitored
B1600+434 with the VLA in A- and B-arrays at 8.5\,GHz, during a period
of about 8 months in 1998 (Koopmans et al. 2000).  From these light
curves we determined a time delay of 47$^{+5}_{-6}$ days (68\%), using
the minimum-dispersion method from Pelt et al. (1996). The PRH method
(Press, Rybicki \& Hewitt 1992) gives the same time delay within the
error range. Shifting the light curve of image B back in time over
this time-delay and subsequently subtracting it from the light curve
of image A, taking the proper flux-density ratio in to account,
results in the difference light curve shown in Fig.2.  The rms
variability of the difference light curve (2.8\%) is significantly
larger than the expected rms variability due to noise only
(1.1\%). The difference light curve is consistent with the presence of
non-intrinsic variability (i.e. `external' variability) at the
14.6--$\sigma$ confidence level (KdB00). Because the time delay was
determined using a minimum-dispersion method, time delays other than
47 days would yield an even larger rms variability in the difference
light curve.

\begin{figure}[t!]
\begin{center}
  \leavevmode
\vbox{%
  \epsfxsize=\hsize
  \epsffile{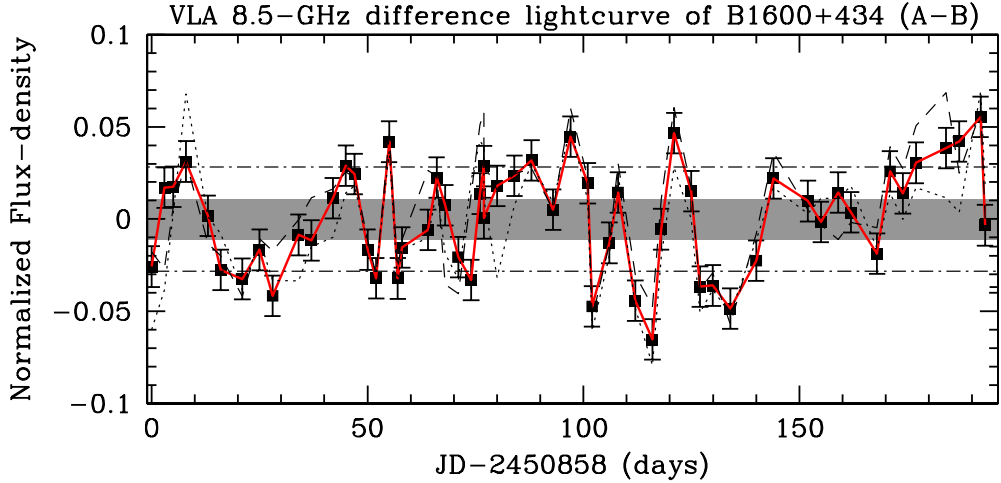}
}
\end{center}
\caption{The normalized difference light curve between the two lensed
images, corrected for both the time-delay and flux density ratio
(Sect.2.1).  The shaded region indicates the expected 1--$\sigma$
(1.1\%) region if all variability were due to measurement errors. The
dash--doted lines indicate the observed modulation-index of 2.8\%. The
dotted and dashed curves indicate the normalized difference curves for
a time delay of 41 and 52 days (68\% confidence region), respectively.}
\end{figure}

\section{`External' variability in B1600+434}

What can be the origin of the presence of `external' variability in 
the image light curves? There are several plausible causes:
\begin{itemize}

\item Scintillation by the Galactic ionized interstellar medium.

\item Microlensing of the background source by massive compact objects
      in the lens galaxy.

\end{itemize}
\noindent
Before proceeding, however, let us first summarize the results on
B1600+434 from our VLA \& WSRT monitoring data (until late 1999), 
that might be relevant in uncovering the true nature of the 
observed external variability (see KdB00):

\begin{enumerate}

\item The short-term rms variability (i.e. the 1$^{\rm st}$--order
modulation index) of the lensed images A and B are 2.8\% and 1.6\%
respectively.

\item The difference light curve has an rms variability of 2.8\%.

\item The VLA 8.5-GHz image light curves, as well as the difference
light curve, seem to show long-term ($\gg$1\,d) variability, next to
some faster variability. The precise time scales are hard to
determine, because of the average light-curve sampling rate of 
once per 3.3 days.

\item Several 5-GHz WSRT 12-h integrated flux-density lightcurves of
B1600+434 (June-July 1999), with a sampling rate of once every 5 min, 
show no evidence for variability $\ga$2\% on time scales less than 
12 hours.

\item Multi-frequency monitoring with the WSRT shows a decrease in 
the short-term rms variability from 5 to 1.4\,GHz by a factor of
$\sim$3.

\end{enumerate}

\noindent
Can this considerable body of multi-frequency data be explained 
in terms of Galactic scintillation (Section 4.1) or radio-microlensing
(Section 4.2)?

\subsection{Galactic Scintillation}

From `standard' scintillation theory (e.g. Narayan 1992), we expect a strong
anti-correlation between the time-scale of variability and its
amplitude (i.e. rms variability). According to the Taylor \& Cordes
(1993) model for the Galactic ionized medium, B1600+434 should be in the
weak-scattering regime and have a variability time scale of at most
$\sim$1 day for an observed rms variability of 1.6--2.8\%.  However,
the WSRT data show no evidence for this short-term variability (point
4). In fact a number of flux-density variations appear to have
relatively long time scales (several weeks; see point 3 and also
Section 5).  These time scales would normally correspond to an rms
variability that is well below the noise level ($\ll$1\%).  Hence, 
the longer variability time scales are {\sl not} compatible
with the observed amplitude of variability in the lensed images (KdB00).

Beside this, the images are separated by only 1.4 arcsec but show a
factor $\sim$1.75 difference in their rms variabilities (point 1).
This either results from a difference of a factor $\sim$3.1 in their
Galactic scattering measures, or a 90--$\mu$as scatter-broadening of
image B in the lens galaxy at $8.5\times(1+z_{\rm s})$$\approx$12\,GHz
(KdB00).  This requires a scattering measure $\sim$1~kpc\,m$^{-20/3}$ 
in the lens galaxy, which is very large compared to typical lines-of-sight 
through our Galaxy. Scatter-broadening at a lower level can not be excluded,
however.

Finally, in the case of scintillation the observed rms variability
should increase towards longer wavelenghts if the source is extended
(i.e. larger than the scattering disk). In the case of B1600+434,
however, we find that the rms variability decrease by almost a factor
$\sim$3 between 5 and 1.4\,GHz (KdB00; point 5).  
One might argue that the source
contains a compact component that scintillates {\sl and} a more extended
component that does not vary. The compact component would show less
variability at 1.4\,Ghz than at 5\,GHz, because it would be in the
strong-scattering regime. However, the time-scale of variability at 5\,GHz
for this component would remain only a few hours, whereas the 12-h
WSRT light curves show no evidence for this (point 4).

Overall, we conclude that the observational evidence does not seem to
favor scintillation as the explanation of the observed external
variability in B1600+434 (KdB00). To save this hypothesis a considerable
rethinking of the properties of the Galactic ionised medium is
required, at least in the direction of B1600+434.

\subsection{Radio-microlensing}

How about radio-microlensing? There are different `ingredients' that play
a role: 
\begin{itemize}

\item{\bf Source Structure:} We assume that the source consist of a
stationary core with a jet structure. This jet contains (several)
condensations, either physical `bullets' or shock-fronts. These 
jet-components move with near- or super-luminal motion and are 
compact ($\mu$as scale), especially near the core. Components
further along the jet might grow in size.

\item{\bf Compact Objects in the Lens Galaxy:} The lens galaxy
consists (partly) of massive compact objects. These objects create 
a magnification pattern on the source-plane, if they constitute
a considerable fraction of the lens-galaxy mass inside a 
radius of 0.5--1 arcsec,

\end{itemize}
\noindent
When these two `ingredients' are combined, one will have
jet-components that move over the magnfication pattern,
being continuously magnified and/or demagnified, resulting in
variability in the integrated flux density of the lensed
images.  The time-scale for these
jet-component to cross the Einstein diameter of a point-mass is
$$\Delta t \sim 4\,(1+z_{\rm s})\,\beta^{-1}_{\rm app}
\cdot\sqrt{(M/M_\odot)}\sqrt{(D_{\rm ds}D_{\rm s}/ D_{\rm d}{\rm
~Gpc})} \mbox{~~~weeks,}$$ where $z_{\rm s}$ is the source redshift,
$\beta_{\rm app}$ is the apparent jet-component velocity in units of
$c$, $M$ is the lens mass, and $D_{\rm ds}$, $D_{\rm s}$ and $D_{\rm
d}$ are the angular diameter distances between lens-source,
observer-source and observer lens, respectively.  More complex
microlensing simulations show similar time-scales of the order of
weeks to months for typical jet velocities of a few time $c$ and
compact-object masses of $\sim$1\,M$_\odot$. The longer observed
variability time-scales for B1600+434 are therefore consistent with
the radio-microlensing hypothesis (KdB00).

The difference in rms variability between the lensed images can be
explained either by a difference in the compact-object mass function
in their respective lines-of-sight and/or by moderate
scatter-broadening ($\ga$few\,$\mu$as) of the image passing through
the disk/bulge.  If for example the average mass of compact objects in
the disk/bulge is much lower than that for the halo line-of-sight, the
resulting magnification pattern will show a more dense caustic
network. This reduces the expected rms variability for a given angular
source size (e.g. KdB00). 

The strongest argument in favor of microlensing, however, is the
frequency dependence of the observed rms variability. Using the
constraints on the jet-structure that we derived from the 8.5-GHz VLA
observations in 1998, we predict a decrease in modulation index from 5
to 1.4\,GHz by a factor $\sim$3, if we assume that the jet-components
are synchrotron self-absorbed (i.e. grow linearly with
wavelength). This decrease has indeed been observed in the 
independent WSRT observations from 1998/1999 (Section 4; KdB00).

\section{A caustic crossing in the radio?}

During the 1999/2000 VLA A- and B-array configurations we have been
monitoring B1600+434 at 8.5, 5 and 1.4\,GHz (Koopmans et al. in prep).
Some very preliminary results at 5\,GHz are shown in Fig.3. This data
show several strong `events' (10--30\%) at both 5 and 8.5\,GHz,
whereas they are absent in the 1.4-GHz light curve.  What is their
origin? All distinct events in image A do not re-occur in image B
(after the time-delay of $\sim$47 days) and are therefore {\sl not}
intrinsic source variability.

Can the strongest feature (around day 80) be an `extreme scattering
event' (ESE; Fiedler et al. 1987)? Plasma clouds in our Galaxy move
with typically $\sim$30 km/s. Event durations of several weeks then
imply an angular size for these clouds of $\sim$1\,mas, if they are
located at $\sim$0.5\,kpc distance from us, thereby covering the entire
source. An event of $\sim$30\% at 5\,GHz would then be detectable at
1.4\,GHz as well, whereas no evidence for this is seen in the VLA
1.4-GHz light curves.  Moreover, these very rare ESE's are not expect
to occur several times in the same light curve of image A within a
time span of only $\sim$150 days. Similarly, also scintillation does not
show this extreme behavior in amplitude over time scales
of several weeks (see Section 4.1).

Rather, we think that this event is caused by a $\mu$as-scale
jet-component in the lensed source, which has recently been ejected
from the core. This component subsequently moves over a single 
caustic in the magnification pattern that is created by compact 
objects in the line-of-sight towards lensed image A, causing 
observable variability in the integrated flux-density of the
lensed image.

\begin{figure*}[t!]
\parbox[b]{7.2cm}{
\caption{Preliminary results (at 5\,GHz) from the 1999/2000 VLA
monitoring campaign of B1600+434. The upper light curve (image A)
passes through the dark-matter halo of the edge-on spiral lens galaxy
(Fig.1). Note several strong (up to 30\%) events (solid arrows) in 
the upper lightcurve and the complete absence of these events 
in the lower light curve (image B) after the time delay of 
$\sim$47 days (dashed arrows).}\vspace{0.4cm}}
\hfill
\resizebox{6.5cm}{!}{\includegraphics{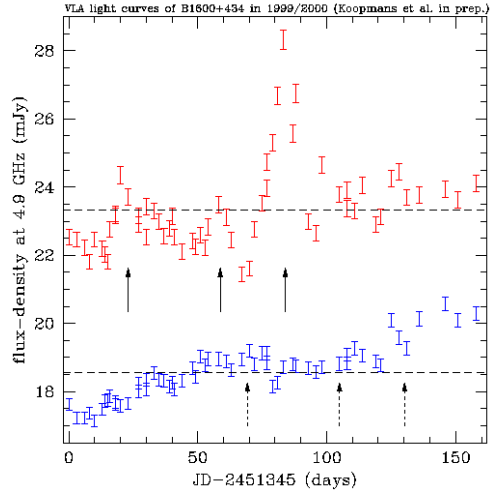}}
\end{figure*}

\section{Conclusions}

The CLASS gravitational lens B1600+434 shows strong `external'
variability in the VLA 5 and 8.5-GHz light curves of lensed image A,
which passes predominantly through the halo of the lens galaxy.
Neither Galactic scintillation nor Extreme Scattering Events (ESE) can
satisfactorily explain all the observations. Only `radio microlensing'
can explain the observed variability time-scales and amplitudes, as
well as its frequency dependence, without invoking extreme
assumptions. It also offers a natural explanation for the strong
($\sim$30\%) event that we more recently observed, as being a
`radio-microlensing caustic crossing'. 
 
Based on the 1998 VLA and 1998/9 WSRT data, we already attempted to
place constraints on the mass-function of compact objects in the
lens-galaxy halo, finding a lower limit of $\ga$0.5\,M$_\odot$
(KdB00).  However, these constraints are still weak and depend
strongly on assumptions about scatter-broadening in image B.  The way
foreward is to use {\sl only} the lightcurve of image A to put
constraints on the mass function in its line-of-sight.  With the
detection of distinct radio-microlensing events, this task has become
a realistic possibility. We have started to work on this (Koopmans et
al. in prep) and hopefully we will soon be able to place significantly
better constraints on the mass function of compact objects in the
lens-galaxy halo, bringing cosmological radio-microlensing up to
Galactic standards!

\acknowledgments

We thank Roger Blandford, Konrad Kuijken, Jean-Pierre Marquart, Penny
Sackett and Jane Dennett-Thorpe for critical discussions during the
course of this work. We also thank the participants of the
`Microlensing 2000' conference for positive feedback. LVEK and AGdeB
acknowledge the support from an NWO program subsidy (grant number
781-76-101). This research was supported in part by the European
Commission, TMR Program, Research Network Contract ERBFMRXCT96-0034
`CERES'.

\end{document}